# Anisotropic Black Phosphorus Synaptic Device for Neuromorphic Applications

He Tian[1,‡], Qiushi Guo[2,‡], Yunjun Xie[3], Huan Zhao[1], Cheng Li[2], Judy J. Cha[3], Fengnian Xia[2,*], and Han Wang[1,*]

[1]*Ming Hsieh Department of Electrical Engineering, University of Southern California, Los Angeles, CA 90089*

[2]*Department of Electrical Engineering, Yale University, New Haven, Connecticut 06511*

[3]*Department of Mechanical Engineering and Materials Science, Yale University, New Haven, Connecticut 06511*

‡ These authors contributed equally to this work.

*Corresponding authors: han.wang.4@usc.edu, fengnian.xia@yale.edu

**Synapses are functional links between neurons, through which "information" flows in the neural network. These connections vary significantly in strength, typically resulting from the intrinsic heterogeneity in their chemical and biological properties. Such heterogeneity is fundamental to the diversity of neural activities, which together with other features of the brain enables functions ranging from perception and recognition, to memory and reasoning. Realizing such heterogeneity in synaptic electronics is critical towards building artificial neural network with the potential for achieving the level of complexity in biological systems. However, such intrinsic heterogeneity has been very challenging to realize in current synaptic devices. Here, we demonstrate the first black phosphorus (BP) synaptic device, which offers intrinsic anisotropy in its synaptic characteristics directly resulting from its low crystal symmetry. The charge transfer between the 2-nm native oxide of BP and the BP channel is utilized to achieve the synaptic behavior. Key features of biological synapses such as long-term plasticity with heterogeneity, including long-term potentiation/depression and spike-timing-dependent plasticity, are mimicked. With the anisotropic BP synaptic devices, we also realize a

**simple compact heterogeneous axon-multi-synapses network. This demonstration represents an important step towards introducing intrinsic heterogeneity to artificial neuromorphic systems.**

Biological synapses vary significantly in their connection strengths[1, 2, 3], which is one of the key features enabling the immense diversity in the activities and functionalities of biological neural networks. The synaptic connection strengths are typically quantified in terms of the synaptic weight change[4, 5, 6, 7], i.e. the ratio between the change in post-synaptic-current after a synaptic pulse is applied (ΔPSC) and the initial post-synaptic current (PSC), which is the critical device characteristics enabling potentiation, depression and spike-timing dependent plasticity (STDP). In fact, the term "intrinsic heterogeneity" here refers to the heterogeneity in synaptic device characteristics arising from intrinsic material properties, resembling that in biological synapses due to the variety in their intrinsic acceptor properties in the post-synaptic terminal. Resistive random-access memory (RRAM)[8, 9, 10] and field-effect-transistor (FET)[11, 12, 13, 14] are the two major types of artificial synaptic devices that can mimic synaptic behaviors for brain-inspired computing and storage. However, in the existing RRAM and FET based synaptic devices, the change in the device size alone cannot alter the synaptic weight change since both the initial PSC and ΔPSC can vary proportionally with device size. Moreover, it is technologically challenging to introduce such heterogeneity in RRAM and FET based artificial synaptic devices by means of controlling local doping variation, or by selecting different oxide thickness and material composition for each individual device across a large wafer.

Recently, layered black phosphorus (BP) has attracted significant attention from the electronic and optical device community[15, 16, 17, 18] due to its high carrier mobility, a moderate bandgap, and most interestingly a strong in-plane anisotropy[19, 20, 21] in its electronic properties. Furthermore, its native phosphorus oxide ($PO_x$)[22, 23] typically formed together with BP provides a natural oxide/semiconductor heterostructure ideal for building synaptic devices. In this work, we demonstrate an anisotropic synaptic device based on the $PO_x$/BP heterostructure, in which the synaptic characteristics are realized utilizing the trapping of electrons in the phosphorus oxide. The intrinsic anisotropy in the electronic properties of the BP layer offers heterogeneity in the synaptic characteristics of BP devices constructed along different crystal orientations of the BP layer. The BP synaptic devices also demonstrated long-term plasticity with time constant on the order of hundreds of seconds, which is important for mimicking long-term memory or learning functions in the biological synapses[24, 25, 26]. Furthermore, a simple axon-multi-synapses network with intrinsic heterogeneity is demonstrated, suggesting the potential of BP for developing low-power high-density heterogeneous axon-synaptic network.

The schematic structure of the artificial BP synaptic device is shown in Figure 1a. BP flakes with thickness around 20 nm are typically selected. Raman and infrared spectra characterizations are then performed to identify the in-plane crystal orientation[19, 21] of the BP samples before device fabrication. An important innovation in the device structure is the utilization of BP native oxide $PO_x$ as the key functional layer, in which the transport of the oxygen ions results in the synaptic behavior of the device. In the 15 devices we have fabricated, cumulative exposure of ~30 minutes in ambient environment during flake exfoliation and metal deposition always results in a layer of $PO_x$ with thickness around 2 nm on the bottom side of the BP flake. The fabricated devices were immediately loaded into the vacuum chamber of a Lakeshore probe-station where both the room temperature and high temperature electrical characterizations were performed. The samples were stored in a light-proof nitrogen box and the bottom PO$_x$ thickness does not change up to at least a few months.

In typical operations, the input pulses (emulating the signals from the pre-synaptic neurons) are applied to the gate. The PO$_x$ layer below the BP layer serves as the electrons trapping layer, which can trigger positive or negative change of the current in the BP channel. A novel feature in the BP devices compared to conventional synaptic devices is that both the positive and negative synaptic weight changes are significantly different for devices built along the x- (armchair) and y- (zigzag) directions of the BP crystal. Figure 1b shows the ambipolar transfer characteristics of the x- and y-direction devices, respectively. First, the overall drain current in the x-direction device is higher than that in the y-direction device due to the higher mobility of carriers in the x-direction of the BP crystal.[19] The devices also show clear hysteresis that is essential for synaptic device operation. There are two main effects[27] that can cause hysteresis in the transfer characteristics of a low-dimensional transistor device: (1) the charge transfer between the interfacial layer and the channel and (2) the charge in interfacial layer inducing opposite charge in channel, i.e. the capacitive gating. Typically, charge transfer can cause a positive shift in the minimum conductivity point in backward (from positive to negative) gate seeping if compared with forward sweeping, because electrons are transferred from the channel to interfacial layer in forward gate bias sweeping. The capacitive gating typically causes a negative shift, because the local electrical field is enhanced by dipoles in dielectric and attract more electrons in forward sweep, which need more negative voltage to reach the minimum conductivity point in the backward sweep. In Figure 1b, it is clear that the minimum conductivity point shifts positively in the backward sweep, so the hysteresis in the transfer characteristics are attributed to charge transfer, which is further confirmed in our synaptic property measurements.

The cross-section of the device and the composition of the $PO_x$ layer are characterized by aberration-corrected scanning transmission electron microscopy (STEM). The annular dark-field STEM image in Figure 1c clearly shows the highly crystalline layered lattice of BP, which has a thickness of ~14 nm with 27 layers. The BP layer-to-layer spacing is, hence, around 0.53 nm, which agrees well with the previously reported interlayer distance in BP of 0.53 nm[19]. It is also clear from the STEM image that there is a 2 nm amorphous layer between the $SiO_2$ and BP, and another similar 4 nm amorphous layer above the BP. To identify the nature of this amorphous layer, high resolution energy-dispersive X-ray spectroscopy (EDX) mapping is utilized to investigate the composition of the corresponding regions in the TEM image. As shown in Figure 1c, the crystalline BP layers show strong presence of phosphorus (P) element only, while the $SiO_2$ region shows the strong presence of oxygen (O) element, but no phosphorus. In the interface region (the areas within the white dotted lines), both P and O elements are present, confirming that the interfacial layers are phosphorus oxides on both sides of black phosphorus regions. The top phosphorus oxide layer is thicker than the bottom layer due to its direct exposure to the environment during the device fabrication. The EDX line-scan performed along the cross-sectional region further confirms the presence of $PO_x$ and its thickness. As shown in Figure 1d, the profiles of P, O and Si elements along the cross-section indicate a clear transition region where both P and O are present. Hence, the STEM and EDX results clearly confirm the presence and the thickness of the $PO_x$ and BP layers. Furthermore, the electron energy loss spectroscopy (EELS) measurement performed on the cross-section of the device (see Supporting Information Figure S1) reveals peak at 130 eV for the BP layer and peak at 137 eV for the POx layer.

Figure 2a shows the positive synaptic responses to a positive input pulse (20 V amplitude, 10

ms duration) applied at the gate of the BP devices built along x- (upper panel) and y-directions (lower panel) of the BP crystal, respectively. Previously, the electrons trapping due to the applied electric field has been observed in a variety of native oxides.[28, 29, 30, 31] Hence, a positive pulse applied at the gate can induce positive PSC in the BP channel. Furthermore, it is clear that despite similar increases in PSC (84 nA in the x-direction device, 88 nA in the y-direction device), the synaptic weight change (ΔPSC/PSC) is very different in the x- and y-direction devices. As shown in Figure 2a, the PSC change along the y-direction (11%) is 2.7 times higher than that along the x-direction (4.1%), for devices with identical dimensions. This can be explained by the different change in the carrier mobility along x- and y-directions (see detailed analysis in the Supplementary Information) after the positive pulse is applied and electrons being trapped in the $PO_x$. The reduction in carrier mobility is mainly due to increased scattering of the channel carriers by the charges in the $PO_x$ layer. A more significant relative degradation in the carrier mobility along x-direction is observed compared to that along the y-direction of the BP crystal (see Supplementary Information and Figure S2). As a result, the overall percentage increase in PSC is higher for devices built along the y-direction of the BP crystal compared to that along the x-direction. This anisotropic synaptic response is consistently observed in all (> 15) devices we have measured. Moreover, the thickness dependence of BP synaptic performance is discussed and shown in Figure S3. This intrinsic anisotropy can be used as a new tuning parameter for emulating the variable connection strengths in biological synapses.

After six seconds delay following the application of the positive gate pulse, a negative gate pulse is applied (-20 V amplitude, 10 ms duration) to both the devices along the x- and y-directions. The negative gate pulse releases the trapped electrons from the $PO_x$ layer. The removal of the trapped electrons leads to a shift of the minimum conduction point in the reverse direction, resulting in a lower PSC in the BP channel, and hence a negative post-synaptic response. As a result, the negative pulse generates smaller PSC compared to the initial value of PSC. As shown in Figure 2c, the decrease in PSC is 10 nA (0.5%) and 25 nA (3%) in the devices built along the x- and y-directions, respectively. Here, we would like to point out that the negative change in the relative PSC only occurs if the positive pulse has been applied prior to the negative pulse, i.e. the electrons needs to be injected into the PO$_x$ layer first before it can be removed. It resembles the behavior of biological synapses, which need to be excited first (enhancement of PSC) before they can be inhibited (suppression of PSC). In addition, the negative synaptic weight change is smaller than the positive weight change for devices along both x- and y-directions of the BP crystal, indicating that the negative pulse can only partially remove the electrons injected into the PO$_x$ layer during the previous equal-magnitude positive pulse.

To further investigate the synaptic behavior and the role played by the charges in the PO$_x$ layer, temperature-dependent measurements for the PSC decay after the application of a positive pulse is carried out (Figure 2c). Temperature dependent measurement of the current decay can provide information about the electron de-trapping process in the PO$_x$ layer. As shown in Figure 2c, when the temperature increases from 300 K to 390 K in steps of 30 K, the relaxation time decreases since at higher temperatures, the electrons are more likely to overcome the ionization activation energy.[32, 33] The PSC decay time $\tau$ is related to the temperature and the ionization activation energy as[32, 33]:

$$A + \ln\left(\frac{1}{\tau(T)}\right) = \frac{E}{k\mathrm{T}} \qquad (1)$$

where $E$ is the ionization activation energy; $k$ is the Boltzmann constant; $T$ is the temperature; and $A$ is a fitting parameter.

By plotting $\ln(1/\tau(T))$ as a function of 1000/$T$, the activation energy can be extracted. As shown in Figure 2d, the experimental data is well fitted by a straight line. The ionization activation energy of 0.16 eV for removing the electrons from the $PO_x$ layer can be extracted from the slope of the line. Moreover, PSC decay time constant is on the order of hundreds of seconds. This decay of the PSC magnitude with long time scale, resulting from gradual removal of the electrons, closely resembles the long-term potentiation (LTP)[34, 35, 36] responsible for long-term memory and learning in biological synapse, which also has typical time scale of hundreds of seconds or longer. This LTP behavior is attributed to the relatively long lifetime of the trapped charges in $PO_x$, which is very different from previously reported ion gel[11] or nano-granular $SiO_2$[13] dielectrics with short-term potentiation.

Using these BP devices, the synaptic potentiation and depression can be mimicked. Figure 3a and 3b show the potentiation and depression characteristics of two BP synaptic devices built along the x- and y-direction of the BP crystal, respectively. It is clear that in BP synaptic devices, the potentiation and depression behaviors are very different for devices built along different orientations of the BP crystal. As shown in Figure 3a and 3b, 20 positive pulses (20 V amplitude, 10 ms duration, and 250 ms intervals between pulses) are first applied to the BP synapse followed by 20 negative pulses (-20 V amplitude, 10 ms duration, and 250 ms intervals between pulses) in this measurement. The PSC increases more rapidly in the first few positive pulse cycles and then slightly saturates. When the negative pulse-train is subsequently applied, the PSC is gradually decreased. It is clear that the change in relative PSC is much more significant in the device built along the y-direction compared to the device built along the x-direction of the BP crystal.

In biological synapse, the spike timing dependent plasticity (STDP)[37] is important for the learning and memory functions of the brain. The STDP of BP synaptic devices built along the

x- and y-directions of the BP crystal are shown in Figure 3c and 3d, respectively. When the pre-synaptic spike arrives before the post-synaptic action, it results in strengthening of the synaptic connection (potentiation). Longer positive time interval gives weaker potentiation. On the contrary, it leads to weakening of the synaptic connection (depression) when the pre-synaptic spike arrives after the post-synaptic spike. Longer negative time interval reduces the depression response of the device. The measured behavior of the BP synaptic devices agrees well with the STDP in biological systems[38]. An exponential relation[8, 39] can be fitted to the data to extract the time constants for both potentiation and depression responses[38], which is based on a spike-timing dependent synaptic plasticity rule in which the amount of synaptic modification arising from a single pair of pre- and postsynaptic spikes separated by a time Δt is given by[38]:

$$G(\Delta t) = \begin{cases} A_+ \exp\left(\frac{\Delta t}{\tau_+}\right) & \text{if } \Delta t < 0 \\ -A_- \exp\left(\frac{\Delta t}{\tau_-}\right) & \text{if } \Delta t \geq 0 \end{cases} \quad (2)$$

The range of pre-to-postsynaptic inter-spike intervals over which the strengthening and weakening of synaptic connections occur is given by τ+ and τ-. They are both in the range of tens of milliseconds, which match well with the millisecond-scale response in typical biological systems[40]. Moreover, consistent anisotropy in STDP and its time constants are observed in all devices we measured. The synaptic weight change in the y-direction device is larger than that in the x-direction device. The time constants τ+ and τ- are smaller for the y-direction device compared to the x-direction device. There is significantly larger weight changes in the y-direction device compared to the x-direction device for both 10 ms or -10 ms inter-spike intervals, and the weight change decays to a similar level at 50 ms or -50 ms for both devices. The weight change in the y-direction device is hence more rapid, resulting in a smaller time constant in the y-direction device.

The anisotropic synaptic behavior in the BP devices is ideal for mimicking the different connection strengths among biological synapses. Both the schematic device structure and the fabricated device are shown in Figure 4a. In biological neural network, multiple synapses with heterogeneous connection strengths typically branch out from a single axon[41] (Figure 4b). Here, we demonstrate a compact realization of an artificial axon-multi-synapses network with heterogeneous synaptic connection strengths using anisotropic BP synaptic devices. Eight electrodes equally spaced at 45 degree angle with respect to each other are fabricated on the BP sample, leading to four diagonal pairs of electrodes with BP channel along different crystalline orientations modulated by a common back gate. The resulting compact device structure closely resembles that of an axon linked with four synapses, each having different connection strengths due to the different orientations of the synaptic devices relative to the BP crystal. Figure 4c shows the connections from the axon to the multiple synapses network and then to the dendrites and neurons. The red box outlines the axon-multi-synapses network that is emulated by the compact BP device shown in Figure 4a. An input pulse train applied to the gate electrode can lead to simultaneous, yet anisotropic, PSC responses in all the four BP synaptic devices. Figure 4d shows the PSC signals simultaneously measured in two out of the four synaptic devices built along the x- and y-directions subject to positive (upper panel) and negative (lower panel) pre-synaptic pulse trains, respectively. Each input pulse has a waveform resembling typical pulses[40, 42, 43] in biological synaptic network. The pulse train in the upper panel in Figure 4c has sequential voltage levels of -15V, 25V, -1V, -1V, -1V, -1V, -6V for a total duration of 130 ms. The interval between neighboring pulses is 8 s. The signal polarity is reversed for the pulse train in the lower panel of Figure 4c. The two synaptic devices can operate simultaneously in response to both positive and negative pulse train applied at the gate (axon) electrode. There is a clear difference in the synaptic weight (ΔPSC/initial PSC) between the synaptic devices built along different directions of the BP crystal for both positive and negative (Figure 4e) responses, which emulates an axon-multi-synapse network with intrinsic connection heterogeneity. The weight change is significantly larger in the y-direction synaptic device compared to that in the x-direction device for both positive and negative responses (Figure 4e). In addition, the plots in Figure 4d clearly show again the long-term plasticity of both the x- and y-direction BP synaptic devices. The enhanced PSC in response to positive pulses and the suppressed current in response to negative pulses both show negligible change during the 8 s intervals between the pulses. This highly compact heterogeneous axon-multi-synapse network represents a promising initial step for introducing heterogeneity into artificial synaptic network, leveraging the intrinsic in-plane anisotropy in black phosphorus.

In summary, we realized the first black phosphorus based synaptic transistor, which represents the first artificial synapses leveraging the crystal anisotropy of the channel material to emulate heterogeneity in biological synapses. In addition, these devices utilize the native oxide $PO_x$ of BP as interfacial layer, giving rise to the synaptic behavior through the trapping and de-trapping of electrons in response to positive and negative pulses applied to the gate. Finally, an artificial anisotropic axon-multi-synapses network is demonstrated using a highly compact BP synaptic structure, emulating a biological axon-multi-synapse network with connection heterogeneity. With the additional tuning of synaptic device behavior enabled by the anisotropy of the BP channel, BP based synaptic devices are promising for a new generation of neuromorphic electronics to emulate the complex heterogeneity in biological neural network.

**Methods**

***Device Fabrication:*** BP flakes were first exfoliated from bulk crystals by micromechanical exfoliation method onto 90 nm $SiO_2$ on Si substrate, which is then spin-coated with PMMA. Electron beam lithography is then used to define the electrode patterns. 3 nm/40 nm Cr/Au were deposited by electron beam evaporation followed by lift-off to form electrical contact with BP. After the device fabrication, the device is characterized in the vacuum chamber of the Lakeshore cryogenic probe station.

*AFM Measurements:* The AFM images were captured using a Bruker Dimension-Icon FastScan system.

*Transmission Electron Microscopy:* To prepare a cross-section sample using a focused ion beam (FIB), a sputtered layer of ~40 nm Si was first coated onto the samples' surface using a K575X Emitech coating system. The samples were then placed in an FEI Helios dual beam FIB/SEM system. An additional layer of ~60 nm C was deposited by injection of an organo-metallic gas and rastering the electron beam over the area of interests. Thin cross sections were extracted from the die surface using an in-situ FIB technique. The cross sections were attached to a TEM grid using FIB-deposited platinum. One window in each section was thinned to electron transparency using the gallium ion beam of the FEI FIB. The STEM images of the sample were then obtained using an aberration-corrected Hitachi HD2700 scanning transmission electron microscope with 200 kV acceleration voltage.

*Energy-dispersive X-ray spectroscopy:* Elemental data was acquired using a Bruker Quantax EDX system.

*Electrical Measurements:* All the electrical characterizations were carried out using Agilent B1500A parameter analyzer in a Lakeshore cryogenic probe station. The electrical pulses with 10 ms duration were applied to the gate as pre-synapse input and the behaviors of the drain current were recorded as the post-synapse signals. Temperature dependent measurements from 300 K to 390 K were performed in the Lakeshore probe station with substrate heating through the chuck. All measurements were performed in vacuum ($<1\times10^{-4}$

Torr).

**Competing financial interests**

Authors declare no competing financial interests.

**References**

1. Kandel, E. R. The molecular biology of memory storage: a dialogue between genes and synapses. *Science* **294**, 1030-1038 (2001).

2. Huttenlocher, P. R. & De Courten C. The development of synapses in striate cortex of man. *Hum. Neurobiol.* **6**, 1-9 (1986).

3. Buonomano, D. V. & Merzenich, M. M. Cortical plasticity: from synapses to maps. *Annu. Rev. Neurosci.* **21**, 149-186 (1998).

4. Royer, S. & Paré, D. Conservation of total synaptic weight through balanced synaptic depression and potentiation. *Nature* **422**, 518-522 (2003).

5. Castro, C., Silbert, L., McNaughton, B. L. & Barnes, C. Recovery of spatial learning deficits after decay of electrically induced synaptic enhancement in the hippocampus. *Nature* **342**, 545-548 (1989).

6. Yang Y, Chen B, Lu WD. Memristive Physically Evolving Networks Enabling the Emulation of Heterosynaptic Plasticity. *Adv. Mater.,* 10.1002/adma.201503202 (2015).

7. Abbott, L., Varela, J., Sen, K. & Nelson, S. Synaptic depression and cortical gain control. *Science* **275**, 221-224 (1997).

8. Jo, S. H., Chang, T., Ebong, I., Bhadviya, B. B., Mazumder, P. & Lu, W. Nanoscale memristor device as synapse in neuromorphic systems. *Nano Lett.* **10**, 1297-1301 (2010).

9. Ohno, T., Hasegawa, T., Tsuruoka, T., Terabe, K., Gimzewski, J. K. & Aono, M. Short-term plasticity and long-term potentiation mimicked in single inorganic synapses. *Nat. Mater.* **10**, 591-595 (2011).

10. Yu, S., Wu, Y., Jeyasingh, R., Kuzum, D. & Wong, H.-S.P. An electronic synapse device based on metal oxide resistive switching memory for neuromorphic computation. *IEEE Trans. Electron Devices* **58**, 2729-2737 (2011).

11. Kim, K., Chen, C. L., Truong, Q., Shen, A. M. & Chen, Y. A carbon nanotube synapse with dynamic logic and learning. *Adv. Mater.* **25**, 1693-1698 (2013).

12. Shi, J., Ha, S. D., Zhou, Y., Schoofs, F. & Ramanathan, S. A correlated nickelate synaptic transistor. *Nat. Commun.* **4**, 2676 (2013).

13. Zhu, L. Q., Wan, C. J., Guo, L. Q., Shi, Y. & Wan, Q. Artificial synapse network on

inorganic proton conductor for neuromorphic systems. *Nat. Commun.* **5**, 3158 (2014).

14. Tian H., et al. Graphene dynamic synapse with modulatable plasticity. *Nano Lett.*, 10.1021/acs.nanolett.5b03283 (2015).

15. Li L., et al. Black phosphorus field-effect transistors. *Nat. Nanotechnol.* **9**, 372-377 (2014).

16. Ling, X., Wang, H., Huang, S., Xia, F. & Dresselhaus, M. S. The renaissance of black phosphorus. *Proc. Natl. Acad. Sci.* **112**, 4523-4530 (2015).

17. Liu H, *et al.* Phosphorene: an unexplored 2D semiconductor with a high hole mobility. *ACS Nano* **8**, 4033-4041 (2014).

18. Koenig SP, Doganov RA, Schmidt H, Neto AC, Oezyilmaz B. Electric field effect in ultrathin black phosphorus. *Appl. Phys. Lett.* **104**, 103106 (2014).

19. Xia, F., Wang, H. & Jia, Y. Rediscovering black phosphorus as an anisotropic layered material for optoelectronics and electronics. *Nat. Commun.* **5**, 4458 (2014).

20. Qiao, J., Kong, X., Hu, Z.-X., Yang, F. & Ji, W. High-mobility transport anisotropy and linear dichroism in few-layer black phosphorus. *Nat. Commun.* **5**, 4475 (2014).

21. Wang, X., et al. Highly anisotropic and robust excitons in monolayer black phosphorus. *Nat. Nanotechnol.* **10**, 517-521 (2015).

22. Favron, A., et al. Photooxidation and quantum confinement effects in exfoliated black phosphorus. *Nat. Mater.* **14**, 826-832 (2015).

23. Wood, J. D., et al. Effective passivation of exfoliated black phosphorus transistors against ambient degradation. *Nano Lett.* **14**, 6964-6970 (2014).

24. Teyler, T. & DiScenna, P. Long-term potentiation. *Annu. Rev. Neurosci.* **10**, 131-161 (1987).

25. Bliss, T. V. & Collingridge, G. L. A synaptic model of memory: long-term potentiation in the hippocampus. *Nature* **361**, 31-39 (1993).

26. Walsh, D. M., et al. Naturally secreted oligomers of amyloid β protein potently inhibit hippocampal long-term potentiation in vivo. *Nature* **416**, 535-539 (2002).

27. Wang, H., Wu, Y., Cong, C., Shang, J. & Yu, T. Hysteresis of electronic transport in graphene transistors. *ACS Nano* **4**, 7221-7228 (2010).

28. Tian, H., et al. In Situ Tuning of Switching Window in a Gate‐Controlled Bilayer Graphene‐Electrode Resistive Memory Device. *Adv. Mater.*, 10.1002/adma.201503125 (2015).

29. Lv, H., et al. Endurance enhancement of Cu-oxide based resistive switching memory with Al top electrode. *Appl. Phys. Lett.* **94**, 213502 (2009).

30. Nigo, S., et al. Conduction band caused by oxygen vacancies in aluminum oxide for resistance random access memory. *J. Appl. Phys.* **112**, 033711 (2012).

31. Sleiman, A., Sayers, P. & Mabrook, M. Mechanism of resistive switching in Cu/AlOx/W nonvolatile memory structures. *J. Appl. Phys.* **113**, 164506 (2013).

32. Wang, Z. Q., Xu, H. Y., Li, X. H., Yu, H., Liu, Y. C. & Zhu, X. J. Synaptic learning and memory functions achieved using oxygen ion migration/diffusion in an amorphous InGaZnO memristor. *Adv. Funct. Mater.* **22**, 2759-2765 (2012).

33. Nian, Y., Strozier, J., Wu, N., Chen, X. & Ignatiev, A. Evidence for an oxygen diffusion model for the electric pulse induced resistance change effect in transition-metal oxides. *Phys. Rev. Lett.* **98**, 146403 (2007).

34. Gerstner, W., Kistler, W. M., Naud, R. & Paninski, L. Neuronal dynamics: From single neurons to networks and models of cognition. Cambridge University Press (2014).

35. Morris, R., Anderson, E., Lynch, G. & Baudryl, M. Selective impairment of learning and blockade of long-term potentiation by an N-methyl-D-aspartate. *Nature* **319**, 774-776 (1986).

36. Malenka, R. C. & Nicoll, R. A. Long-term potentiation--a decade of progress? *Science* **285**, 1870-1874 (1999).

37. Caporale, N. & Dan, Y. Spike timing-dependent plasticity: a Hebbian learning rule. *Annu. Rev. Neurosci.* **31**, 25-46 (2008).

38. Bi, G.-Q. & Poo, M.-M. Synaptic modifications in cultured hippocampal neurons: dependence on spike timing, synaptic strength, and postsynaptic cell type. *J. Neurosci.* **18**, 10464-10472 (1998).

39. Kim, S., Du, C., Sheridan, P., Ma, W., Choi, S. & Lu, W. D. Experimental Demonstration of a Second-Order Memristor and Its Ability to Biorealistically Implement Synaptic Plasticity. *Nano Lett.* **15**, 2203-2211 (2015).

40. Song, S., Miller, K. D. & Abbott, L. F. Competitive Hebbian learning through spike-timing-dependent synaptic plasticity. *Nat. Neurosci.* **3**, 919-926 (2000).

41. Toni, N., Buchs, P.-A., Nikonenko, I., Bron, C. & Muller, D. LTP promotes formation of multiple spine synapses between a single axon terminal and a dendrite. *Nature* **402**, 421-425 (1999).

42. Markram, H., Lübke, J., Frotscher, M. & Sakmann, B. Regulation of synaptic efficacy by coincidence of postsynaptic APs and EPSPs. *Science* **275**, 213-215 (1997).

43. Gerdeman, G. L., Ronesi, J. & Lovinger, D. M. Postsynaptic endocannabinoid release is critical to long-term depression in the striatum. *Nat. Neurosci.* **5**, 446-451 (2002).

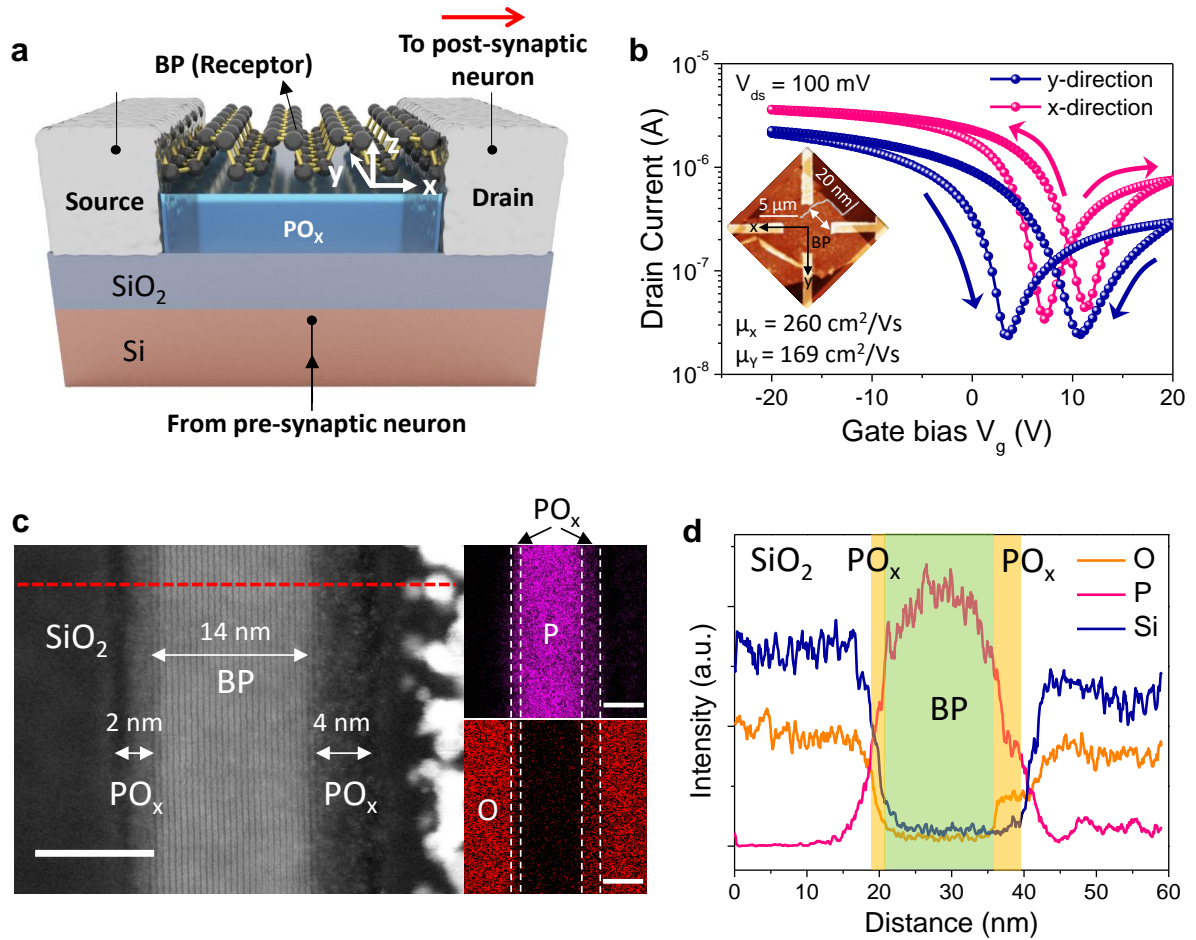

**Figure 1. BP synaptic device and phosphorus oxide characterization.** (a) Schematic of the BP synaptic device. Input pulses are applied to the pre-synaptic terminal, i.e. the silicon back-gate. Output signals are measured at the post-synaptic terminal, i.e. the drain. (b) The transfer characteristics ($I_d$-$V_g$) showing hysteresis due to the charge transfer to the $PO_x$ layer for two BP synaptic devices built along the x- and y-directions of the BP crystal, respectively. Carrier mobility along the x-direction is higher than the mobility along the y-direction. Inset: the AFM image of the BP synaptic device showing devices built along both the x- and y-directions of the BP crystal. The sample thickness is ~20 nm. (c) Scanning Transmission Electron Microscopy (STEM) image showing the cross-section of the device and the corresponding high resolution energy-dispersive X-ray spectroscopy (EDX) mappings of the P and O elements. All scale bars: 10 nm. The $PO_x$ layer between the BP layers and the $SiO_2$

underneath has a thickness of 2 nm. (d) The EDX line-scans showing the profiles of the P, O and Si elements along the cross-section of the device, i.e. along the red dotted line in the STEM image in Figure 1c.

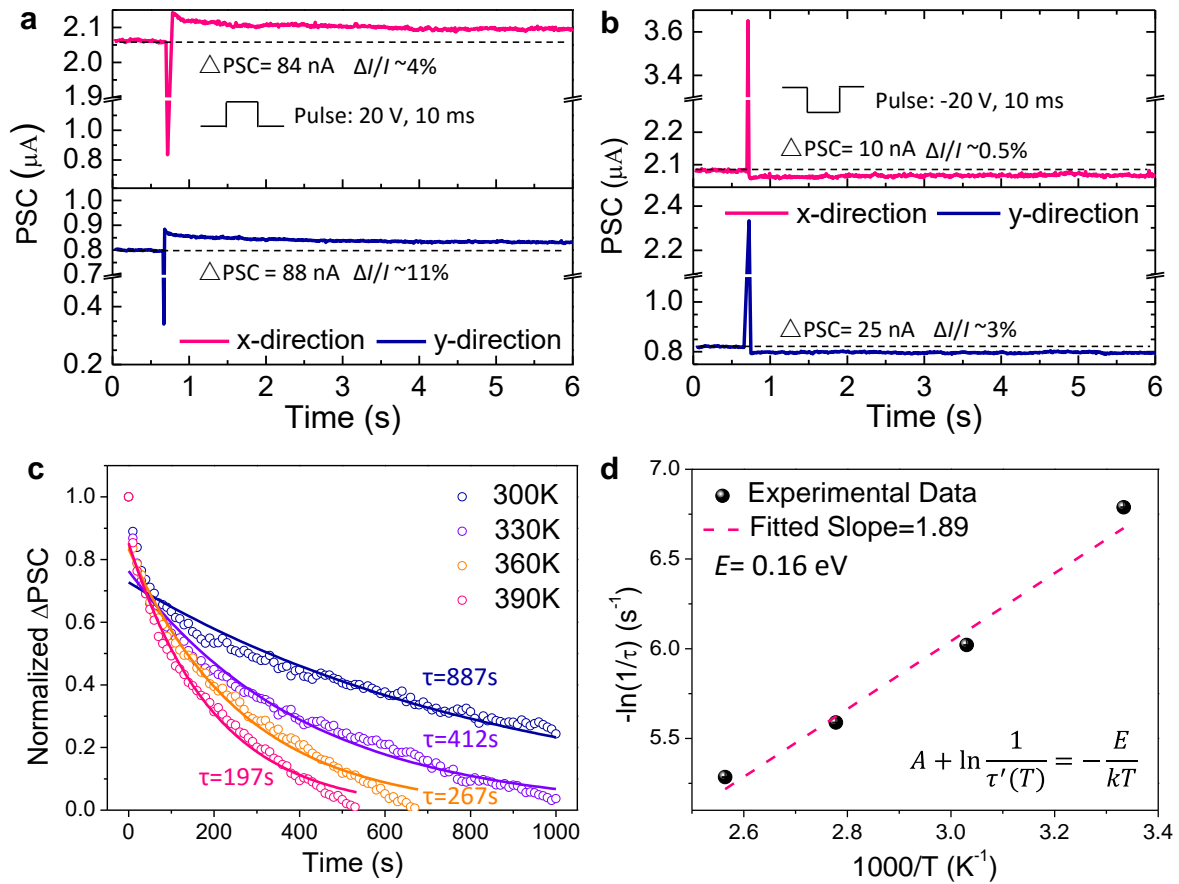

**Figure 2. BP synaptic behavior and temperature dependent measurements.** (a) The positive PSC changes in the BP synaptic devices built along the x- and y-directions of the BP crystal, respectively, in response to a positive pulse at the gate (20 V amplitude, 10 ms duration). (b) The negative PSC changes in the BP synaptic devices built along the x- and y-directions of the BP crystal, respectively, in response to a negative pulse at the gate (-20 V amplitude, 10 ms duration). (c) The temperature dependent measurements characterizing the PSC decay after a positive input pulse is applied. (d) The plot of $-\ln(1/\tau)$ vs. $1000/T$. The ionization activation energy is extracted to be 0.16 eV.

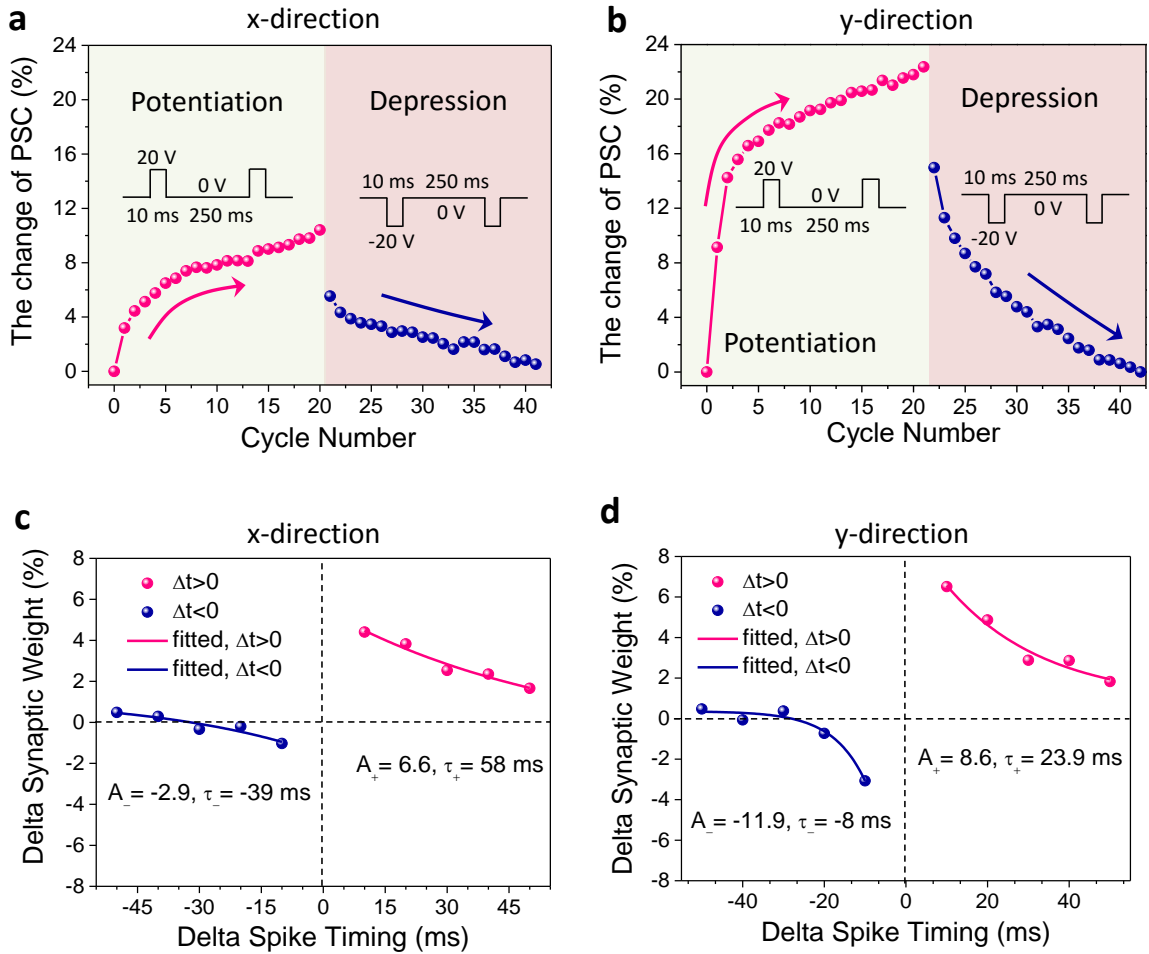

**Figure 3. Anisotropic BP synaptic behavior: potentiation, depression and STDP.** (a) and (b): The PSC modulation of the BP synaptic devices subject to a train of positive (20 V, 10 ms pulses with 250 ms intervals) and negative (-20 V, 10 ms pulses with 250 ms intervals) pulses at the gate for the x-direction and the y-direction devices, respectively. The extent of potentiation and depression in the y-direction device is larger than that in the x-direction device. (c) and (d): Anisotropic STDP characteristics in the x-direction and the y-direction devices, respectively.

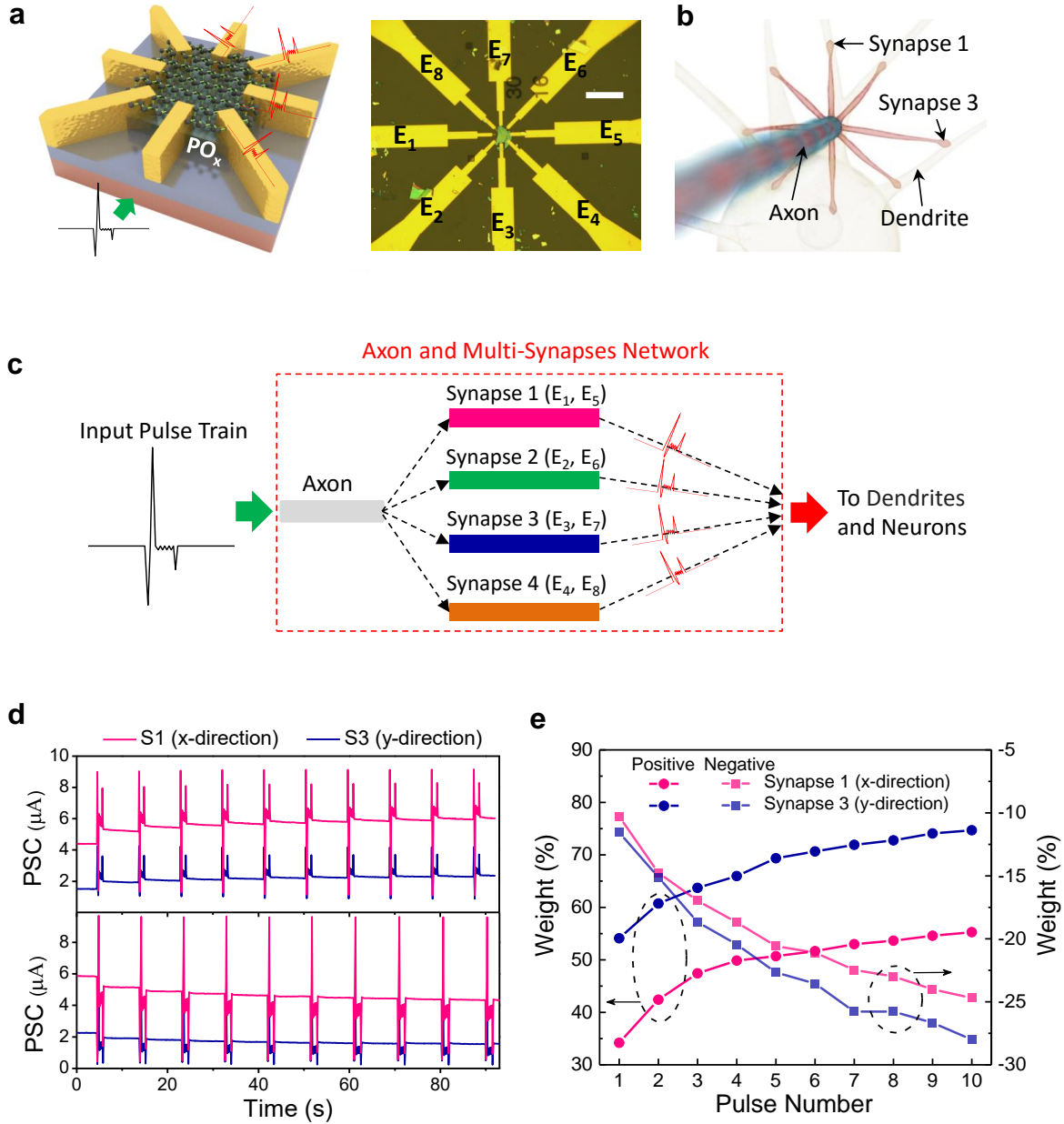

**Figure 4: A compact artificial axon-multi-synapses network with heterogeneous connection strengths.** (a) The schematic (left) and an optical micrograph (right) of a BP synaptic network. Scale bar: 25 μm. (b) A schematic showing a bio-synaptic network. One axon can connect to multiple synapses with different connection strengths. The BP synaptic network in (a) resembles a bio-synaptic network. (c) The device testing scheme. Input pulses are applied to the axon (silicon back gate) and the PSC changes of each synaptic device are recorded from the dendritic terminal (drain). (d) PSC signals simultaneously recorded from Synapse 1 (x-direction) and Synapse 3 (y-direction). The waveform of each input pulse is selected to resemble typical pre-synaptic pulses in biological systems. Each pulse in the upper panel has sequential voltage levels of -15V, 25V, -1V, -1V, -1V, -1V, -6V for a total duration of 130 ms. The interval between neighboring pulses is 8 s. The signal polarity is reversed for the pulse train in the lower panel. (e) The simultaneously recorded positive and negative weight changes of Synapse 1 and Synapse 3.

# Supplementary Information

# Anisotropic Black Phosphorus Synaptic Device for Neuromorphic Applications

*He Tian[1], Qiushi Guo[1], Yujun Xie, Huan Zhao, Cheng Li, Judy J. Cha, Fengnian Xia\*, and Han Wang\**

Dr. He Tian, Mr. Huan Zhao, Prof. Han Wang\*
Ming Hsieh Department of Electrical Engineering, University of Southern California, Los Angeles, CA 90089
\*E-mail: han.wang.4@usc.edu
Mr. Qiushi Guo, Mr. Cheng Li, Prof. Fengnian Xia\*
Department of Electrical Engineering, Yale University, New Haven, Connecticut 06511
\*E-mail: fengnian.xia@yale.edu
Mr. Yujun Xie, Prof. Judy J. Cha
Department of Mechanical Engineering and Materials Science, Yale University, New Haven, Connecticut 06511

([1]These authors contributed equally to this work.)

## I. EELS of BP and PO$_x$

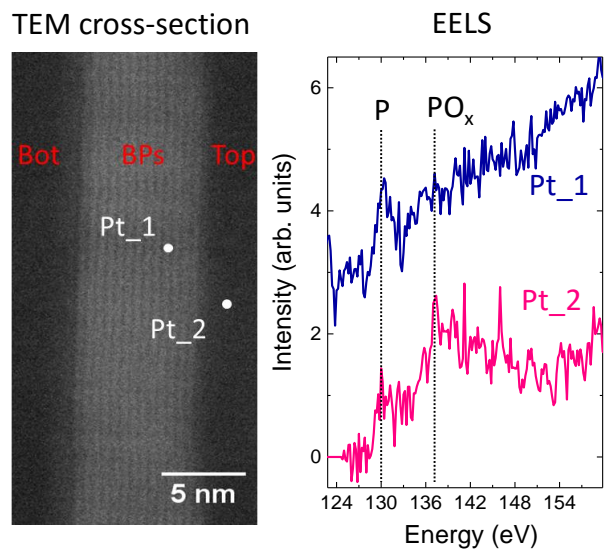

**Figure S1. STEM cross-section and electron energy loss spectroscopy (EELS) for BP and PO$_x$ layers.** As shown in Figure S1, based on the high angle annular dark field STEM image, the middle crystalline region is the BP layer and both the top and bottom regions are amorphous. EELS spectra (~nm spatial resolution) were obtained from the cross-section of the device corresponding to locations Pt_1 and Pt_2 in the STEM image. EELS measurement confirms that the middle layer (Pt_1) is dominated by phosphorus with a strong peak around 130 eV. For the amorphous layer (Pt_2), the peak around 130 eV is much weaker compared to the intensity of the same peak at Pt_1. Moreover, a distinct peak at 137 eV can be found for Pt_2 but not for Pt_1, which corresponds to PO$_x$ and agrees well with the peak position

measured in previous study of PO$_x$ using EELS[S1]. Here, by performing the EELS study on the cross-section of the BP sample for the first time, we are able to clearly distinguish between the crystalline BP layer and the amorphous oxidized layer using the cross-sectional STEM image before explicitly measuring the EELS spectra for the BP (Pt_1) and PO$_x$ (Pt_2) layers respectively. The results provide strong evidence for the presence of the PO$_x$ layer and its composition. For the EELS study, a 200 kV FEI Tecnai Osiris TEM at Yale Institute for Nanoscience and Quantum Engineering was used.

## II. Anisotropic Synaptic Weight Change

Here, we analyze the different weight changes for BP synaptic devices with identical dimensions built along x- and y-directions of the BP crystal. The change in PSC for x- and y-direction devices can be calculated as following:

$$I = (\frac{W}{L}V_{ds}) \cdot \mu \cdot (C_S V_{gs}) \tag{S1}$$

where $W$ is channel width, $L$ is channel length, $V_{ds}$ is the drain bias voltage. $(\frac{W}{L}V_{ds})$ has the same value in both x- and y-direction devices. Here the device channel length is long (> 5 μm) and we ignore the contact resistance for simplicity. The PSC difference between x- and y-direction devices has two major contributions: the difference in carrier mobility μ and the difference in the injection carriers $ne = C_S V_{gs}$. Then eq. (S2) can be written as:

$$I_0 = (\frac{W}{L}V_{ds}e) \cdot \mu \cdot n \tag{S2}$$

After the gate pulse, there is an accumulation of trapped charges in the PO$_x$ layer, which can cause scattering and reduce the carrier mobility in the channel. The PSC after the application of the gate pulse is:

$$I = (\frac{W}{L}V_{ds}e) \cdot (\mu + \Delta\mu) \cdot (n + \Delta n) \tag{S3}$$

Then the PSC change can be written as:

$$\Delta I = I - I_0 \approx \left(\frac{W}{L}V_{ds}e\right) \cdot (\mu \cdot \Delta n + \Delta\mu \cdot n) \tag{S4}$$

$$\Delta I_x - \Delta I_y = \left(\frac{W}{L}V_{ds}e\right) \cdot [(\mu_x - \mu_y)\Delta n + \Delta\mu_x \cdot n_x - \Delta\mu_y \cdot n_y] \tag{S5}$$

Then the synaptic weight change can be written as:

$$W = \frac{\Delta I}{I_0} = \frac{\Delta n}{n} + \frac{\Delta\mu}{\mu} \tag{S6}$$

Hence, the synaptic weight change in x- and y-direction devices are:

$$W_x = \frac{\Delta n_x}{n_{x0}} + \frac{\Delta\mu_x}{\mu_{x0}} \tag{S7}$$

$$W_y = \frac{\Delta n_y}{n_{y0}} + \frac{\Delta\mu_y}{\mu_{y0}} \tag{S8}$$

For the $\frac{\Delta n_x}{n_{x0}}$ and $\frac{\Delta n_y}{n_{y0}}$, the oxygen vacancies and trapped charges induced by the gate pulse are expected to be independent of the device crystalline directions.

$$\frac{\Delta n_x}{n_{x0}} = \frac{\Delta n_y}{n_{y0}} \tag{S9}$$

The main difference should be related to the $\frac{\Delta\mu_x}{\mu_{x0}}$ and $\frac{\Delta\mu_y}{\mu_{y0}}$ difference.

The mobility can be written as:

$$\mu = \frac{\Delta I_{SD}}{\Delta V_{SG}}\left(\frac{L}{W}\frac{1}{V_{ds}C_S}\right) \tag{S10}$$

The mobility μ before and after pulse can be estimated by the slope of transfer curve $\frac{\Delta I_{SD}}{\Delta V_{SG}}$ at zero gate voltage.

After calculate the mobility (See following **Figure S2**), we can obtain

$$\frac{\Delta \mu_x}{\mu_x} = \frac{163-260}{260} = -0.373, \quad \frac{\Delta \mu_y}{\mu_y} = \frac{117-169}{169} = -0.308 \tag{S11}$$

Then you can find the

$$W_y - W_x = \left(\frac{\Delta \mu_y}{\mu_y} + \frac{\Delta n_y}{n_{y0}}\right) - \left(\frac{\Delta \mu_x}{\mu_x} + \frac{\Delta n_x}{n_{x0}}\right) \approx 0.072 = 7.2\% \tag{S12}$$

The intrinsic weight difference from calculation using the I-V characteristics is 7.2%. Our time domain measurements shows the weight difference is 7% (**Figure 2a**). The difference in synaptic weight changes between the x- and y-direction devices is mainly due to the mobility change in x- and y-direction after the pulse. The trapped charges due to the gate pulse in PO$_x$ can suppress the higher mobility in the x-direction more significantly.

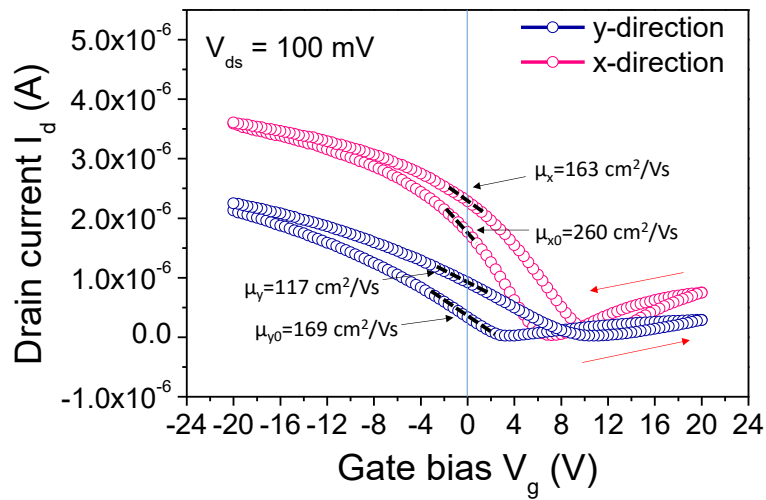

**Figure S2. The transfer curves in linear scale showing the field effect mobility estimated from forward and backward sweeps.**

## III. Thickness Dependence of the BP Synapse

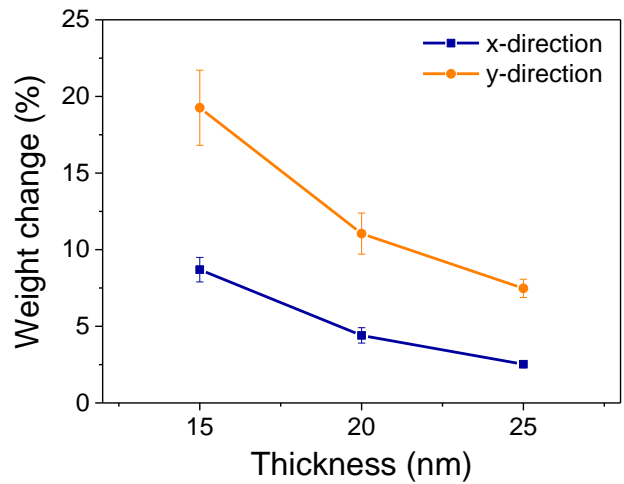

**Figure S3. The weight change vs. total BP film thickness.**

We have performed the synaptic testing on additional BP samples with different thicknesses. As shown in **Figure S3**, all the samples show the larger weight change in y-direction than that of the x-direction. For thicker samples, both the x- and y-direction weight changes are lower. This is because the total charges in BP increases with the thickness and the screen effect is stronger. In fact, the BP transistor on/off ratio also decreases as the thickness increases. As a result, the weight changes are suppressed. When the BP thickness decreases, the weight changes in both the x- and y-directions increase, reflecting the reduced screening effect and more significant carrier modulation in the channel[S2-S4].

**References**

[S1] Favron, Alexandre, et al. "Photooxidation and quantum confinement effects in exfoliated black phosphorus." *Nature Materials* 14.8 (2015): 826-832.

[S2] Yang, Zhibin, et al. "Field-Effect Transistors Based on Amorphous Black Phosphorus Ultrathin Films by Pulsed Laser Deposition." *Advanced Materials* 27.25 (2015): 3748-3754.

[S3] Xia, Fengnian, Han Wang, and Yichen Jia. "Rediscovering black phosphorus as an anisotropic layered material for optoelectronics and electronics." *Nature Communications* 5 (2014): 4458.

[S4] Li, Likai, et al. "Black phosphorus field-effect transistors." *Nature Nanotechnology* 9.5 (2014): 372-377.